\documentclass[sigconf,screen]{acmart}
\copyrightyear{2025}
\acmYear{2025}
\setcopyright{cc}
\setcctype{by}
\acmConference[FSE Companion '25]{33rd ACM International Conference on the Foundations of Software Engineering}{June 23--28, 2025}{Trondheim, Norway}
\acmBooktitle{33rd ACM International Conference on the Foundations of Software Engineering (FSE Companion '25), June 23--28, 2025, Trondheim, Norway}
\acmDOI{10.1145/3696630.3727245}
\acmISBN{979-8-4007-1276-0/2025/06}

\usepackage{hyperref}
\usepackage{url}
\usepackage{enumitem}
\usepackage[skip=2pt]{caption}
\usepackage{multirow}
\usepackage{pgfplots}
\usepackage{colortbl}
\usepackage{xspace}
\usepackage{balance}

\pgfplotstableread[row sep=\\,col sep=&]{
    interval & carT \\
    0 & 46  \\
    1--5 & 61  \\
    6--10 & 7 \\
    11--15 & 1 \\
    16--20 & 7\\
    20+ & 7 \\
    }\mydata

\begin{document}

\title[How do Software Engineering Candidates Prepare for Technical Interviews?]{How do Software Engineering Candidates Prepare for\\Technical Interviews?}

\author{Brian Bell}
\affiliation{%
 \institution{Virginia Tech}
 \city{Blacksburg}
 \state{VA}
 \country{USA}
 }
\email{bbell97@vt.edu}

\author{Teresa Thomas}
\affiliation{
  \institution{Virginia Tech}
  \city{Blacksburg}
  \state{Virginia}
  \country{USA}
}\email{teresa2020@vt.edu}

\author{Sang Won Lee}
\affiliation{
  \institution{Virginia Tech}
  \city{Blacksburg}
  \state{Virginia}
  \country{USA}
}
\email{sangwonlee@vt.edu}

\author{Chris Brown}
\affiliation{
  \institution{Virginia Tech}
  \city{Blacksburg}
  \state{Virginia}
  \country{USA}
}
\email{dcbrown@vt.edu}

\newcommand{\todo}[1]{{\color{red}\bfseries [[TODO: #1]]}}

\newcommand{\ie}{\textit{i.e.,}\xspace}

\newcommand{\etal}{\textit{et al.}\xspace}

\begin{abstract}
To obtain employment, aspiring software engineers must complete \textit{technical interviews}---a hiring process which involves candidates writing code while communicating to an audience. However, the complexities of tech interviews are difficult to prepare for and seldom faced in computing curricula. To this end, we seek to understand how candidates prepare for technical interviews, investigating the effects of preparation methods and the role of education. We distributed a survey to candidates ($n = 131$) actively preparing for technical interviews. Our results suggest candidates rarely train in authentic settings and courses fail to support preparation efforts---leading to stress and unpreparedness. Based on our findings, we provide implications for stakeholders to enhance tech interview preparation for candidates pursuing software engineering roles.
\end{abstract}



\maketitle

\section*{Prelude}

\begin{figure*}[h]
        \includegraphics[width=0.75\textwidth]{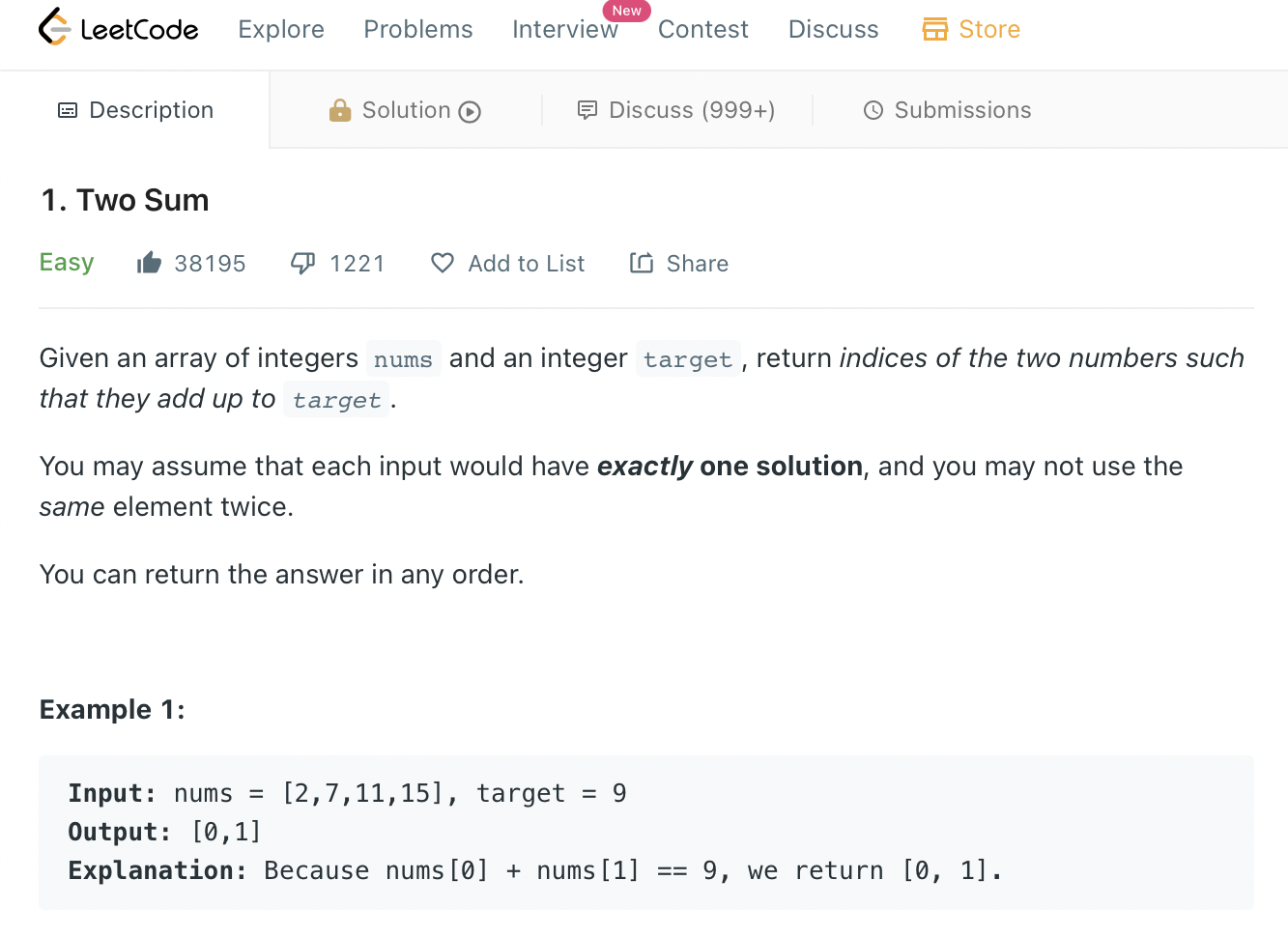}
        \caption{Two Sum technical interview problem, as portrayed by \url{
    https://leetcode.com/problems/two-sum/}}
        \label{fig:Leetcode2sum}
\end{figure*}

Consider the \textit{Two Sum} problem depicted in Figure~\ref{fig:Leetcode2sum}, an ``easy'' challenge on LeetCode~\cite{leetcode}---an online platform for technical interview preparation. Take a moment to try and solve this problem---Were you able to come up with a solution? Does it pass the provided test case? What are the time and space complexity? Can you verbally describe how you came up with your solution? Could you do this while being watched by an interviewer?---According to the LeetCode community, this problem was used in recent tech interviews to hire software engineers at companies such as Google, Meta, Amazon, and LinkedIn.\footnote{\url{https://leetcode.com/discuss/interview-experience?query=two sum}}

\section{Introduction}

Software engineers are primarily hired through an evaluation process known as \textit{technical interviews}. Traditional technical interviews, or tech interviews, consist of candidates writing code to solve a programming challenge related to data structures and algorithms--typically in a non-programming environment such as a whiteboard or online text editor (\ie Google Docs or CollabEdit)--while performing a \textit{think-aloud} to verbally communicate their problem-solving strategy and thought process to observing interviewer(s)~\cite{CrackingCodingInterview,whiteboard}. This process is widely used by companies to evaluate the proficiency of candidates pursuing software engineering (SE) positions, assessing the technical and soft skills needed to work with a team to develop complex and high-quality software systems~\cite{kaatz}. 

However, the experiences of candidates provide a different perspective of current tech hiring practices. Technical interviews invoke negative perceptions and frustrations from developers, who criticize this process for its data structure-focused nature and irrelevance to real-world SE work, among other complaints~\cite{mahnaz2019hiring,mahnaz2020DH}. Further, companies that do not utilize traditional tech interviews in their hiring processes are praised for hiring without whiteboards~\cite{HiringWithoutWhiteboards}. Research also shows technical interviews increase anxiety and lead to worse performances for candidates~\cite{behroozi2020does} and a ``leaky pipeline'' in which qualified candidates are overlooked or burn out in tech hiring processes~\cite{LeakyPipeline}. 


Technical interviews are a cognitively and socially demanding activity---assessing candidates' problem-solving ability, technical knowledge, computational thinking, and communication---making them difficult to prepare for. Recent studies by interviewing.io\footnote{\url{https://interviewing.io}, a tech interview prep website} suggest 54\% of candidates pass interviews~\cite{interviewing100k} and only 20\% perform consistently between interviews~\cite{interviewingio-gap}. Further, aspiring software engineers rarely face authentic technical interview settings---writing code with simultaneous think-aloud in front of observers---in Computer Science (CS) education or in practice, increasing the burden on individuals to prepare~\cite{bell2023understanding}. Yet, a notable complaint from developers is the amount of preparation time and effort needed to be competitive~\cite{cui2024much,mahnaz2019hiring}---with numerous resources recommending candidates spend multiple hours per day practicing coding problems to obtain offers~\cite{harper2022interview,handbook,faang}. Yet, this leads to increased anxiety, particularly for underrepresented job seekers~\cite{hall2018effects,lunn2021uneven}, and bias favoring candidates with more time and resources to prepare~\cite{mahnaz2019hiring}. Moreover, candidates receive no feedback after completing technical interviews, only a job offer or not, making it difficult to improve for future interviews without any skills assessment of past performances~\cite{mahnaz2020DH}. In addition to preparation difficulties, candidates do not learn interview concepts in CS curricula and instructors face challenges teaching tech interview skills to help students succeed~\cite{lunn2024educational}.

In this work, we aim to investigate how candidates \textit{prepare} for the complexities of technical interviews and examine the role of computing education in preparation efforts. To this end, we aim to answer the following research questions (RQs):

\begin{itemize}[noitemsep,topsep=0pt]
    \item[\textbf{RQ1}] How do candidates prepare for technical interviews?
    \item[\textbf{RQ2}] How do candidates perceive the effects of technical interview preparation?
    \item[\textbf{RQ3}] How do candidates perceive the impacts of SE education on preparing for technical interviews?
\end{itemize}

To answer these questions, we distributed an online survey to collect data on candidates' ($n = 131$) preparation resources and methods, the perceived effectiveness of their approaches, and the adequacy of their computing education to support preparation efforts. We found candidates use a variety of preparation resources---yet rarely practice in authentic environments, leading to a lack of preparedness and anxiety for technical interviews. Further, participants suggest computing education is unsupportive for tech interview preparation, expressing a desire for increased support in academic settings. Based on our results, we discuss opportunities to enhance technical interview preparation. Particularly, we provide guidelines for candidates, employers, interview preparation resources, and higher education---focusing on the learning and teaching of technical interview concepts---to enhance candidates' preparedness for complex technical interview environments.

\section{Motivating Example}

Referring back to the ``Two Sum'' problem presented in the Prelude, we provide a brief explanation to demonstrate the difficulties of technical interview preparation. 

First, candidates must implement a solution to the problem in a selected programming language or pseudocode. The brute force algorithm involves looping through each element \textit{x} to find if there is a second element which meets the case of equaling \textit{target - x}. This solution is inefficient with an $O(n^2)$ time complexity, however only uses $O(n)$ space complexity. Thus, candidates must consider more complex data structures and algorithms to provide a more efficient solution.

Another solution is ``\texttt{Two-pass Hash Table}''~\cite{twopass}. This algorithm uses a hash table data structure to reduce the program runtime by mapping array elements to their index. Candidates must recognize this algorithm decreases the time complexity of the solution to $O(n)$ and maintains $O(n)$ space complexity. These are just two of the many solutions that exist for this problem---and Two Sum is one of the innumerable number of problems that could be asked during interviews. In addition to a valid technical solution, candidates must also be prepared to provide a clear verbal rationale describing their approach, which is critical in hiring decisions~\cite{denae2017tech}, in front of one or more interviewers assessing their performance. 


While advanced data structures such as hash tables are commonly taught in CS education, they are complex for students to understand~\cite{mtaho2024difficulties} and infrequently taught across institutions~\cite{simon2010making}---causing students to lack educational training on how to apply these constructs to coding challenges and communicate effectively about them. Moreover, despite being a critical part of software development~\cite{li2015makes}, training in communication and soft skills are often overlooked computing curricula and SE courses~\cite{gonzalez2011teaching}. Consequently, candidates must spend a considerable amount of individual time and effort preparing for technical interviews~\cite{mahnaz2019hiring,cui2024much}.


\section{Related Work}

This project extends prior work investigating technical interviews. Research shows tech interviews are problematic, with developers finding current processes irrelevant, anxiety-inducing, and biased~\cite{mahnaz2019hiring}. Behroozi and colleagues analyzed comments posted by software developers on Glassdoor~\cite{glassdoor}, an online platform for job reviews, and found additional concerns, such as a bias towards candidates with more time and resources to prepare~\cite{mahnaz2020DH}. Ford \etal found mismatches between candidate expectations for tech interviews and hiring manager assessment criteria for hiring decisions~\cite{denae2017tech}. Further, Behroozi \etal used head-mounted eye trackers to show increased levels of stress in candidates during technical interviews, resulting in a negative impact on candidates' performance--particularly those identifying as women~\cite{behroozi2020does,mahnaz2018dazed}. To reduce stress during interviews, asynchronous interview approaches have been proposed and shown to enhance the technical and communication ability of candidates~\cite{mahnaz2022async}. Finally, Lunn \etal show negative tech interviews adversely impact students' computing identity~\cite{lunn2021impact}. 

Research has also investigated technical interview preparation. Behroozi \etal show interview preparation is frustrating for software practitioners~\cite{mahnaz2019hiring}. Cui \etal observed preparation techniques for LeetCode users, highlighting correlations between LeetCode ratings and job offers at top tech companies~\cite{cui2024much}---demonstrating the substantial prep time and effort needed to succeed in tech interviews. Interview preparation has also been explored as a barrier to inclusive computing workforces. For instance, Hall and Gosha outline anxieties in technical interview preparation faced by CS students from Historically Black Colleges and Universities (HBCUs)---who often have less resources than CS departments at other institutions \cite{hall2018effects}. Similarly, Lunn \etal show racial and gender minority students are more likely to have cultural experiences that inhibit their ability to prepare~\cite{lunn2021uneven}. Our study extends this work by examining how candidates prepare for technical interviews---contributing insights on the resources and authenticity of training processes used, their perceived effects on preparedness, and the impact of SE education curricula on technical interview preparation for candidates pursuing SE positions in the tech industry.

\section{Methodology}

The goal of this research is to investigate study patterns candidates use to prepare for technical interviews. We acknowledge other types of SE hiring evaluations exist~\cite{assess}, such as take-home coding assessments or interviews focused solely on behavioral qualities without invoking technical skills. However, the scope of our study targets preparation for \textit{whiteboard-styled technical interviews}---or tech interviews that involve candidates writing code to solve data structures and algorithms-based problems in a non-programming environment while thinking aloud in the presence of one or more interviewers~\cite{whiteboard}. 

\subsection{Online Surveys}

We deployed online surveys to collect data from participants actively preparing for tech interviews to obtain SE-related positions. Our study was approved by our institutional review board (IRB).

\subsubsection{Participant Recruitment}

We used purposive sampling to recruit participants from two main sources. The survey was distributed on department listservs at the authors' institution, targeting students preparing for tech interviews to obtain internship and full-time positions. We also collaborated with Exponent~\cite{exponent} and Pramp~\cite{pramp} to share our survey, allowing us to reach a broader sample. Exponent and Pramp are online platforms to help candidates prepare for technical interviews, providing a variety of features such as coding problems, courses, peer mentoring, and expert coaching. Surveys were placed on the homepage of the respective websites to recruit participants.

\subsubsection{Survey Design}  We distributed three versions of the survey each to our institution, Exponent, and Pramp. The surveys collected data on participants' demographic background, general interviewing experience, preparation techniques and challenges, and the role of university education. The three surveys were identical, except the Exponent and Pramp versions contained an additional section specifically collecting feedback on those systems. The responses to these questions were shared directly with our industry partners, and are not included in our analysis as they do not relate to our research goals. The survey included a variety of short-answer, open-ended, and Likert scale questions to describe their technical interview training methods. The questions were formed based on our RQs and insights from our industry partners, and refined through reviews from experts well-versed in empirical SE research and tech interviews as well as a pilot study with two representative participants from our institution, who are excluded from our data. The surveys are available online.\footnote{\url{https://github.com/code-world-no-blanket/TechInterviewPrep}}
 
\subsection{Data Analysis}

We used a mixed methods approach to analyze the survey data. For Likert scale questions, we leverage the Mann-Whitney U test to analyze responses across groups based on participants' demographic background and preparation practices~\cite{arcuri2011practical}. For continuous data, we divide participants into two groups (high/low) based on the median value~\cite{rforhealth}. For open-ended questions, we used an open coding approach to derive themes from participant responses~\cite{braun2006}. Two researchers independently analyzed and categorized responses, then came together to discuss their findings and resolve disagreements. After this process, the researchers finalized a set of themes based on participant responses. We did not obtain IRB approval to share survey responses, however this data can be made available upon request after receiving the necessary permissions.

\subsection{Participants}

We received a total of 131 responses from a diverse sample of participants actively preparing for technical interviews. For each survey, we received 93 responses from Exponent users, 34 from Pramp users, and four from our university. 104 participants (79.4\%) identified as male while 27 (20.6\%) identified as female. In terms of race/ethnicity, we found 64 participants (48.9\%) identified as Asian, 34 (25.9\%) as White, 17 (13.0\%) as Black or African American, 15 (11.5\%) as Hispanic/Latino, five (3.8\%) as American Indian or Alaskan Native, and 11 (8.4\%) as two or more races. 

We found our study population represents a mixture of candidates pursuing SE careers, including students completing technical interviews to obtain their first job or an internship as well as experienced software engineers in employment transition. 35 participants (26.7\%) self-identified as students--representing freshmen (40\%, $n=14$), juniors (22.9\%, $n=8$), seniors (31.4\%, $n=11$), Masters (20\%, $n=7$), and PhD (5.7\%, $n=2$) students at a wide range of global universities. Most participants ($n = 96$, 73.3\%) identified as SE professionals in industry, representing a variety of companies. Overall, participants averaged nine years of industry experience ($mdn = 7$) across professionals and students. Participants self-reported spending approximately 6.5 hours per week ($mdn = 4$) preparing for technical interviews. Most participants ($n = 108$, 82.4\%) completed at least one technical interview, averaging three per participant ($mdn = 4$). Since our focus is tech interview preparation, we include participants who are preparing but have not yet completed an interview.  

\section{Results}


\subsection{RQ1: Technical Interview Preparation}\label{sec:rq1}

To understand how candidates prepare for technical interviews, we examined the resources participants use to prepare and how often training settings reflect authentic tech interview environments.

\subsubsection{Resources}

\begin{figure*}[t]
\centering
        \includegraphics[width=0.75\linewidth]{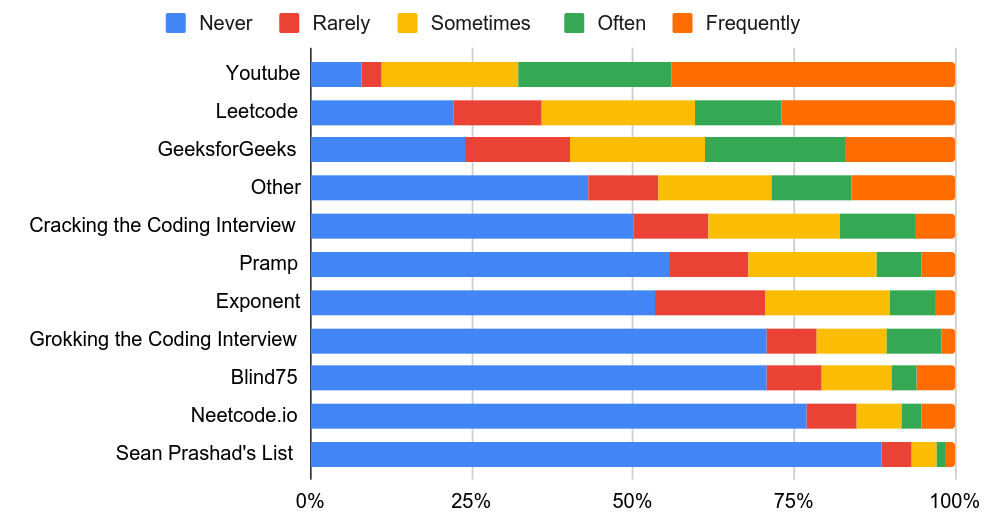}
        \caption{Frequency of reported resources used}
        \label{fig:resource}
\end{figure*}

We found participants use a wide array of technical interview preparation resources are available to help candidates, representing varied mediums with different features. For example, there are online coding practice platforms (\ie LeetCode~\cite{leetcode}--depicted in Figure~\ref{fig:Leetcode2sum}), study guides (\ie Blind75~\cite{blind75}), and physical books (\ie ``\textit{Cracking the Coding Interview}'' ~\cite{CrackingCodingInterview}). YouTube~\cite{youtube}, ($n = 108$), LeetCode~\cite{leetcode} ($n = 90$), and GeeksforGeeks~\cite{geeks} ($n = 77$) were the most popular resources reported and the most frequently used among participants (see Figure~\ref{fig:resource}). Other resources mentioned include Exponent, Pramp, Hackerrank, Udemy, Coursera, Algomonster, educative.io, AlgoExpert, Stack Overflow, and more. Eight participants reported using no resources. We also found candidates are unwilling to pay for tech interview preparation resources, with few participants reporting being likely to subscribe to premium accounts for LeetCode ($n = 28$), Exponent ($n = 12$), and Grokking the Coding Interview ($n = 10$), or purchase the Cracking the Coding Interview book ($n = 30$).

\subsubsection{Authenticity}\label{sec:rq1-auth}

To investigate the authenticity of tech interview preparation, we evaluated the extent to which participants practice: (1) coding; (2) communication; and (3) with observers.

\paragraph{Coding}

Technical interviews are designed to assess the coding abilities and technical knowledge of candidates. Generally, the preparation resources used by participants focus on coding practice for technical interviews. For example, participants who reported using LeetCode spend approximately two hours each day practicing data structures and algorithms-related programming challenges. Moreover, over 85\% of participants rank problem-solving ability as \textsl{Very Important} ($n = 85$) or \textsl{Important} ($n = 27$) when preparing for tech interviews. On average, participants reported specifically practicing LeetCode challenges around two hours per day ($mdn = 1$) and three days per week ($mdn = 2$).

\paragraph{Communication}

Communication is essential in software engineering~\cite{li2015makes}. As such, think-aloud is an integral part of technical interviews~\cite{denae2017tech}, sometimes regarded as more important than coding ability~\cite{LearnToCode}. Our results suggest candidates view communication as critical in technical interviews, with 78\% of participants reporting oral and verbal clarity in interviews as \textsl{Important} ($n = 39$) or \textsl{Very Important} ($n = 63$). However, most participants ($n = 39$) reported only practicing communication skills when preparing for technical interviews \textsl{Occasionally}, with less than half practicing think-aloud \textsl{Sometimes}, \textsl{Often}, or \textsl{Always} (see Table~\ref{tab:communication}).

\begin{table*}
    \centering
        \caption{How often do you practice your communication skills during your technical interview prep session?}
    \begin{tabular}[t]{|c|c|c|c|c|c|c|} \hline
       \textbf{\em Response} & \textbf{Never}  & \textbf{Rarely} & \textbf{Occasionally} & \textbf{Sometimes} & \textbf{Often} & \textbf{Always} \\ \hline
       \textbf{\em n (\%)} & 18 (13.7\%) & 11 (8.4\%) & 39 (29.8\%) & 8 (6.1\%) & 32 (24.4\%) & 23 (17.6\%) \\ \hline
    \end{tabular}
    \label{tab:communication}
\end{table*}

\paragraph{Audience}

\begin{figure*}[h!]
\begin{tikzpicture}
\centering
    \begin{axis}[
            ybar,
            symbolic x coords={0,1--5,6--10,11--15,16--20,20+},
            xtick=data,
        ]
        \addplot table[x=interval,y=carT]{\mydata};
    \end{axis}
\end{tikzpicture}
\caption{Range of responses for the number of mock interviews participants have completed}\label{fig:mock}
\end{figure*}
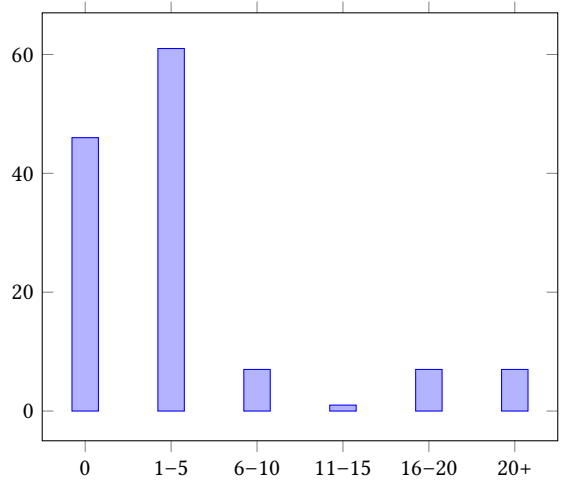

Another key part of technical interview environments is coding and communicating \textit{in front of an observer}. However, we found participants rarely practice with others---as 62\% ($n = 81$) responded \textsl{No} to a question asking if their technical interview preparation includes studying with others. Candidates who reported preparing with others ($n = 50$) mentioned various approaches, including training with friends, family members, colleagues, and study groups in-person and remotely using platforms such as Discord and Zoom. We further investigated the types of preparation with others. Candidates can practice with experts---for example, many preparation resources offer features for users to receive coaching from experts. Yet, participants rarely prepare with tutors (29\%, $n = 38$) and fewer pay for coaching functionalities (31.6\%, $n = 12$). We also explored participants' experience with mock interviews, or simulated technical interviews with peers. However, mock interviews are infrequently used ($avg = 4.6$, $mdn = 2$)---with most respondents (82\%, $n = 107$) completing five or less and 35\% ($n = 46$) never attempting a mock interview (see Figure~\ref{fig:mock}).

\begin{center}

\fcolorbox{black}{yellow!30}{

\parbox{0.96\linewidth}{

\textbf{Finding 1:} Most participants use free, online resources (\ie YouTube, LeetCode, and GeeksforGeeks) to prepare for technical interviews.

\textbf{Finding 2:} We found most candidates do \textit{not} prepare for technical interviews in authentic settings---prioritizing coding practice but rarely preparing in front of an audience with observers.
}
}

\end{center}

\subsection{RQ2: Effects of Tech Interview Preparation}\label{sec:rq2}




To analyze candidates' perceptions of interview preparation, we investigated their self-reported anxiety during interview training and preparedness for interviews.

\begin{table*}
        \centering
        \caption{Causes of Anxiety in Tech Interview Preparation}
        \begin{tabular}{ lr } \toprule
         \textbf{Response} & \textbf{\textit{Participants (n)}} \\ \midrule
            Too many topics to study & 73  \\
            Unsure on what will be asked & 73  \\ 
            Balancing of prep and other time commitments & 69 \\ 
            Poor performance in past interviews & 52  \\ 
            College did not prepare me & 43  \\
            Other & 70 \\
            None & 8 \\ 
            \bottomrule
        \end{tabular}\\
        \small{* Total exceeds number of participants because multiple reasons could be provided in a single response.}
        \label{tab:anxiety}
        \end{table*}

\subsubsection{Anxiety}

We found most participants ($n = 77$, 58.8\%) reported feeling anxious during tech interview preparation. Our qualitative analysis uncovered numerous reasons participants felt anxious while training for interviews, presented in Table~\ref{tab:anxiety}). The main causes were candidates having to study too many topics and the uncertainty of what interviewers will ask. For example, P82 noted ``\textit{It’s mostly the daunting task of trying to remember certain data structures and what you have to recollect and learn again because it is not part of your day to day operations}'' and P131 added ``\textit{There's just so much interviewers could ask you and I simply do not know all of it}''. Another source of anxiety was candidates not having enough time to prepare given their other commitments. For instance, participants noted difficulties regarding ``\textit{the amount of time required to study for interview and the busy working day as I worked as a full-time}'' (P8) and ``\textit{Balancing the amount to time you need to invest to be able to get to a good level with school and jobs and personal time is very hard}'' (P129). The effects of this anxiety also negatively impacted candidates, with participants commenting on preparation, it ``\textit{makes me more anxious before the real interview}'' (P8), ``\textit{I think I'm never prepared or good enough}'' (P23), ``\textit{I have panic attacks all the time and feel like I don't have enough time to prepare}'' (P53), and ``\textit{When I was preparing, I always felt guilty when I wasn't working to prepare more. I felt like I could never take a break}''. These responses suggest the tech interview preparation techniques reported by participants fail to support candidates actively seeking SE positions, resulting in anxiety, frustration, and burnout from their training efforts.

We further investigated whether participants demographic group and preparation methods impacted perceived anxiety. For background, we compare the majority and minority groups for gender, race and status, while dividing number of interviews completed based on the median. For preparation methods, we observed number of reported study \textbf{hours} per week, number of daily \textbf{LeetCode} problems, reported \textbf{communication} practice frequency---based on frequent (\textsl{Always, Often, Sometimes}) and infrequent (\textsl{Occasionally, Rarely, Never}) Likert responses, and whether or not participants reported practicing \textbf{with others}. These results are presented in Table~\ref{tab:rq2_anxiety}. Using a Mann-Whitney U test ($U$), we did not find a significant correlation. This is a potential indicator that the anxiety induced by technical interview preparation impacts all candidates, regardless of background or processes used to prepare.

\begin{table}[t]
\small
\centering
\caption{
Perceived Anxiety Based on Participant Background and Preparation
}
\begin{tabular}{|l|l|r|r|c|} \hline
  \textbf{Background} & \textbf{Type} & \textbf{Median} & \textbf{\textit{p}-value} \\ \hline
 \multirow{2}{*}{\em Gender} & Female &  4  &   $p = 0.2148$  \\ 
 & Male & 4 & ($U = 1246$, $\eta^2 = 0.14$) \\
 \hline
 \multirow{2}{*}{\em Race}
 & non-White & 3 &  $p = 0.3121$  \\
 & White & 4 & ($U =  1329.5$, $\eta^2 = 0.47$) \\
  \hline 
 \multirow{2}{*}{\em Status} 
 & Student  & 4 & $p = 0.3632$   \\
 & non-Student  &  4 & ($U = 1798$, $\eta^2 = 0.06$) \\ \hline
 \multirow{2}{*}{\em Interviews}
 & Lower Half & 4 &  $p = 0.1075$  \\
 & Upper Half & 3 & ($U = 1846.5$, $\eta^2 = 0.01$) \\
 \hline 
 \textbf{Preparation} & \textbf{Type} & \textbf{Median} & \textbf{\textit{p}-value} \\ \hline
 \multirow{2}{*}{\em Hours} & Lower Half &  4  &   $p = 0.3446$  \\ 
 & Upper Half & 4 & ($U = 2027$, $\eta^2 = 0.001$) \\
 \hline
 \multirow{2}{*}{\em LeetCode} 
 & Lower Half  & 4 & $p = 0.2327$   \\
 & Upper Half  &  4 & ($U = 2785.5$, $\eta^2 = 0.08$) \\ \hline
 \multirow{2}{*}{\em Communication} 
 & Frequent  & 4 & $p = 0.3897$   \\
 & Infrequent  &  4 & ($U = 2051.5$, $\eta^2 = 0.001$) \\ \hline
 \multirow{2}{*}{\em With Others} 
 & Yes  & 4 & $p = 0.2033$   \\
 & No  &  4 & ($U = 1850$, $\eta^2 = 0.15$) \\ \hline
\end{tabular}
\label{tab:rq2_anxiety} \\
\textbf{*} denotes statistically significant results (\textbf{p-value < 0.05}), $\eta^2$ is effect size
\end{table}

\begin{table}[t]
\small
\centering
\caption{
Perceived Preparedness Based on Participant Background and Preparation
}
\begin{tabular}{|l|l|r|r|c|} \hline
  \textbf{Background} & \textbf{Type} & \textbf{Median} & \textbf{\textit{p}-value} \\ \hline
 \multirow{2}{*}{\em \textbf{Gender*}} & Female &  3  &   \textbf{$p = 0.0091$*}  \\ 
 & Male & 3 & ($U = 989$, $\eta^2 = 0.04$) \\
 \hline
 \multirow{2}{*}{\em Race}
 & non-White & 3 &  $p = 0.3121$  \\
 & White & 3 & ($U = 1509$, $\eta^2 = 0.004$) \\
  \hline 
 \multirow{2}{*}{\em \textbf{Status*}} 
 & Student  & 4 & \textbf{$p = 0.0418$}*   \\
 & non-Student  &  4 & ($U = 1743$, $\eta^2 = 0.02$) \\ \hline
 \multirow{2}{*}{\em Interviews}
 & Lower Half & 3 &  $p = 0.0838$  \\
 & Upper Half & 3 & ($U = 1815$, $\eta^2 = 0.02$) \\
 \hline 
 \textbf{Preparation} & \textbf{Type} & \textbf{Median} & \textbf{\textit{p}-value} \\ \hline
 \multirow{2}{*}{\em Hours} & Lower Half &  3  &   $p = 0.1587$  \\ 
 & Upper Half & 3 & ($U = 1898$, $\eta^2 = 0.01$) \\
 \hline
 \multirow{2}{*}{\em LeetCode} 
 & Lower Half  & 3 & \textbf{$p = 0.3520$*}   \\
 & Upper Half  & 3 & ($U = 2029.5$, $\eta^2 = 0.001$) \\ \hline
 \multirow{2}{*}{\em \textbf{Communication*}} 
 & Frequent  & 3 & \textbf{$p = 0.0222$*}   \\
 & Infrequent  & 3 & ($U = 1680$, $\eta^2 = 0.03$) \\ \hline
 \multirow{2}{*}{\em \textbf{With Others*}} 
 & Yes  & 3 & \textbf{$p = 0.0256$*}   \\
 & No  &  3 & ($U = 1613$, $\eta^2 = 0.35$) \\ \hline
\end{tabular}
\label{tab:rq2_prep} \\
\textbf{*} denotes statistically significant results (\textbf{p-value < 0.05}), $\eta^2$ is effect size
\end{table}

\subsubsection{Preparedness}

The results for participants' self-reported preparedness are presented in Table~\ref{tab:prepared}. Overall, less than 40\% ($n = 52$) of participants responded positively to being prepared for interviews, with the majority feeling uncertain about their preparedness. 

To further investigate perceived preparedness, we used a Mann-Whitney U to statistically analyze preparedness against participant background and preparation techniques (see Table~\ref{tab:rq2_prep}). For demographic background, we observed a statistically significant difference in perceived preparedness based on the gender of participants (male/female, $p = 0.0091$). This suggests female candidates feel significantly less prepared than male-identifying counterparts while preparing for technical interviews. We also observed experience plays a role in candidates' preparedness, as current software engineers reported feeling significantly more prepared for interviews compared to students ($p = 0.0418$).

\begin{table*}[t]
    \centering
    \caption{Participant Responses to ``\textit{I feel that I am well prepared for the day of my technical interview}''}
    \begin{tabular}{|c|c|c|c|c|c|}\hline
         \textbf{\em Response} & \textbf{Strongly Disagree} & \textbf{Disagree} & \textbf{Neutral} & \textbf{Agree} & \textbf{Strongly Agree} \\ \hline
        \textbf{\em n (\%)} & 7 (5.3\%) & 29 (22.1\%) & 43 (32.8\%) & 38 (29.0\%) & 14 (10.7\%) \\ \hline
    \end{tabular}
    \label{tab:prepared}
\end{table*}

\begin{table*}[h]
    \centering
    \caption{Participant Responses to whether University Courses Prepared them for Technical Interview Concepts}
    \begin{tabular}{|l|r|r|r|r|r|}\hline
           & \textbf{Strongly Disagree} & \textbf{Disagree} & \textbf{Neutral} & \textbf{Agree} & \textbf{Strongly Agree} \\ \hline
        Data Structures & 31 & 31 & 37 & 31 & 14 \\ \hline
        Algorithms & 30 & 27 & 41 & 27 & 13 \\ \hline
        Time/Space Complexity & 24 & 30 & 39 & 30 & 19 \\ \hline
    \end{tabular}
    \label{tab:uni}
\end{table*}

\begin{table*}[h]
    \centering
    \caption{Participant Responses to ``\textit{I would have benefited from a structured study plan provided by my university}''}
    \begin{tabular}{|c|c|c|c|c|c|}\hline
         \textbf{\em Response} & \textbf{Strongly Disagree} & \textbf{Disagree} & \textbf{Neutral} & \textbf{Agree} & \textbf{Strongly Agree} \\ \hline
        \textbf{\em n (\%)} & 14 (10.7\%) & 12 (9.2\%) & 25 (19.1\%) & 30 (22.9\%) & 50 (38.2\%) \\ \hline
    \end{tabular}
    \label{tab:plan}
\end{table*}

We also explored whether various practices impact perceived preparedness. We observed more frequently rehearsing communication ($p = 0.0222$) and practicing with others ($p = 0.0256$) were significant indicators in participants' perceived preparedness---suggesting that authentic practice environments, particularly training communication skills and with others, can improve candidates' perceived preparedness for tech interviews. In addition, contrary to prior work and popular trends in technical interview preparation~\cite{cui2024much,faang,interviewing100k}, we observed that candidates reporting more study hours per week and completed LeetCode problems per day did \textit{not} perceive to be more prepared for technical interviews.



\begin{center}

\fcolorbox{black}{yellow!30}{

\parbox{0.96\linewidth}{

\textbf{Finding 3:} We observed technical interview preparation invokes anxiety for candidates, primarily due to extensive number of concepts, uncertainty of topics, and lack of time to prepare.

\textbf{Finding 4:} Most candidates feel their preparation methods do \textit{not} prepare them for technical interviews. However, candidates who reported practicing communication more frequently as well as training with others reported significantly higher perceived preparedness---while those who reported completing more daily LeetCode problems and studying more hours per week did not.
}
}

\end{center}

\subsection{RQ3: SE Education}\label{sec:rq3}

We asked participants to provide insights on how their university curriculum impacted their tech interview preparation to investigate gaps and highlight competence needs for students pursuing SE roles. When asked whether undergraduate courses helped prepare for interviews, most participants ($n = 64$, 48.9\%) responded \textsl{No}. We further asked participants to rank how well their courses prepared them for concepts common in technical interviews~\cite{CrackingCodingInterview}, shown in Table~\ref{tab:uni}. Most participants were negative or neutral on whether courses prepared them for data structures (75.6\%), algorithms (74.8\%), and time/space complexity of programs (71.1\%). The most common courses that were mentioned as being helpful include Data Structures, Algorithms, Computer Architecture, capstone course, and more--with one participant stating ``\textit{AP Computer Science in High School. It was much more helpful than all the courses I took in college}'' (P45). Further, only one participant mentioned an SE course was useful for technical interview preparation.

We also asked participants if tech interview-related courses would aid preparation efforts---with most participants ($n = 98$, 74.8\%) responding a specific tech interview course would be helpful, while 20 (15.3\%) and 13 (9.9\%) selected \textsl{No} and \textsl{Unsure}, respectively. We also asked if a structured study plan provided by institutions could improve skills needed for technical interviews. The majority of participants ($n = 50$, 38.2\%) strongly agreed that a university or department study plan would support their technical interview preparation efforts (see Table~\ref{tab:plan}). These results provide insights into methods to enhance computing education to better support technical interview preparation for candidates.


\begin{center}

\fcolorbox{black}{yellow!30}{

\parbox{0.96\linewidth}{

\textbf{Finding 5:} Most candidate report university-level CS courses did \textit{not}  adequately prepare them for technical interview concepts, such as data structures and algorithms.

\textbf{Finding 6:} Most candidates believe structured study guides provided by universities can benefit technical interview preparation efforts.
}
}

\end{center}

\section{Discussion}\label{sec:guide}

Our results suggest candidates prepare for technical interviews using a variety of different resources, yet rarely practice in authentic environments with coding, communication, and observers (RQ1). Consequently, most participants feel their preparation leaves them anxious and unprepared for interviews (RQ2). Participants also reported a lack support from their computing education (RQ3). Based on our results, we provide guidelines for how candidates, employers, tech interview preparation resources, and higher education can overcome difficulties in technical interview preparation.

\subsection{For Candidates}\label{sec:guide-cand}

Our findings show that candidates frequently practice coding skills and often rehearse communication during technical interview preparation. However, participants rarely prepared with other people. Beyond increasing the amount of time and effort to prepare~\cite{cui2024much}, which our results suggest may not impact preparedness for interviews (see Table~\ref{tab:rq2_prep} in Section~\ref{sec:rq2}), we posit preparation with others can help candidates enhance learning for interview skills and feel better prepared for hiring assessments. 

\subsubsection{Practice with Others}
Our results show that candidates who prepare with others feel significantly more prepared for their technical interviews than those who practice alone. The public nature of technical interviews has been shown to increase stress and worsen interview performance~\cite{behroozi2020does}. Prior studies also show participants exhibit less stress in interview environments without an observer present~\cite{mahnaz2020DH,mahnaz2022async}. However, interviewer presence is common in interviews~\cite{whiteboard}. Thus, preparing with other people, using approaches such as mock interviews, can enhance preparation for actual interview performances. For example, mock interviews, which were underutilized by candidates in our study, can simulate tech interviews with time pressure, a live observer, and think-aloud communication. 

Mock interviews are also a form of \textit{collaborative learning}, or an educational approach where students work together in groups to learn and solve problems~\cite{laal2012collaborative}, which has been shown to enhance outcomes and learning for students in a variety of domains~\cite{laal2012benefits} and SE contexts~\cite{coccoli2011computer,clarke2014integrating}. For instance, collaborative learning through mock interviews can preparation beyond individual practicing through pooled knowledge, explaining concepts, error-correction, reduced memory load, and observational learning~\cite{nokes2015better}. Thus, practicing with others can provide more relevant experience with authentic environments for technical interviews during  preparation---increasing students' learning and confidence in skills necessary to succeed in technical interviews.

\subsection{For Employers}

While improving technical interview preparation will help candidates, it can also benefit companies. Hiring processes are time-consuming and expensive~\cite{HiddenCost}. To that end, we provide implications for companies to help candidates better prepare and motivate improvements to SE hiring practices. 

\subsubsection{Rethink Technical Evaluations}
We found technical interview preparation anxiety-inducing, regardless of candidate experience. Further, increased preparation efforts do not enhance candidates' perceived preparedness for assessments. Thus, modifications to SE hiring processes are needed to focus candidates' preparation, providing better evaluations of candidates' abilities and their suitability for a role at the company. For example, think-aloud allows employers to gain insight into the thought processes of candidates---however, this is frequently combined with evaluating soft skills~\cite{mahnaz2022async}. We also found candidates infrequently practice communication skills during interview prep, and even more rarely practice in front of an observing audience (see Section~\ref{sec:rq1-auth}). For instance, P25 noted the unfairness of this evaluation approach, stating employers ``\textit{just expected [them] to talk it out like you work there...How is that apple to apples?}''. P100 also noted they lack ``\textit{confidence in putting thoughts to words}''. Thus, separating think-aloud and soft skills assessments can better assess candidates' fit based on their experiences, character, personality traits, aptitude, and communication ability. 

Another concern is the irrelevance of technical interviews to real-world SE work~\cite{mahnaz2019hiring}. For instance, in our survey P4 noted that tech interview prep is ``\textit{a different skill set to real work and you need to be able to memorize solutions which you won't encounter in the real world}''. As such, candidate preparation often does not prioritize learning skills necessary to succeed as a software engineer~\cite{li2015makes}, but focuses on learning concepts needed to pass the interview. P45 highlights this frustration, stating, ``\textit{sadly, many LeetCode questions suffer from you-either-know-it-or-you-dont problem...If you don't know the algorithm, you won't be able to come up with an optimal solution. It has taken geniuses years to come with up with a solution, so how could you be expected to come up with it in 30 minutes?}''. This can be problematic for companies, as ``LeetCoders'', or candidates who memorize solutions to whiteboard-style technical interview problems without necessarily understanding the implementation, can deceive employers due to good tech interview performances~\cite{LeetCoders}. However, these candidates may not perform well in a real-world development environment. In addition, these false positive hires can cause employers to miss out on qualified candidates who may not have performed as well on the coding challenge during the interview. Thus, providing more relevant assessments can improve candidates' preparation and evaluation---thereby improving the quality of software and the overall tech workforce.

\subsubsection{Provide Feedback on Interview Performance}

Another challenge in tech interview preparation is the lack of feedback~\cite{mahnaz2020DH}. Candidates typically do not receive feedback on interview performances, making it difficult to improve for future interviews. Companies avoid giving feedback for various reasons, such as policies and legal implications~\cite{nofeedback} and a lack of infrastructure to provide feedback~\cite{lu2024helping}. However, feedback is critical for enhancing learning~\cite{roberts1996feedback}. For instance, research shows the act of processing feedback leads to better performance~\cite{luft2014learning}. In addition, a lack of feedback has been shown to inhibit learning and cause students to become discontented~\cite{brown2007feedback}. In our survey, we saw a lack of feedback from technical interviews led to anxiety and frustration for participants, such as one who noted ``\textit{Saving time for interview prep then getting rejected is exhausting and disappointing}'' (P115) and another stating ``\textit{I cannot figure out what the interviewer is looking for}'' (P23). 

Other forms of feedback---such as comments on code reviews---has shown to increase learning in SE settings~\cite{bacchelli2013expectations}. Automatically generated feedback has also shown promise in increasing student learning and programming behaviors~\cite{edwards2003rethinking}. Recently, tools have employed large language models (LLMs) to automatically generate feedback on mock interview performances~\cite{career_io}. However, research shows that candidates distrust LLMs in hiring contexts~\cite{vaishampayan2023procedural}, in addition to technical interview settings~\cite{swanand2025chase}. Further, specific feedback from interviewers can help candidates gain insights into the expectations of employers~\cite{denae2017tech}, providing insights on ways to improve their interview performance and lead to potential job offers  the same company or a different organization.

\subsection{Technical Interview Preparation Resources}\label{sec:guide-tools}

Our survey results show candidates use a wide variety of resources for technical interview preparation (see Figure~\ref{fig:resource}). As the goal of these systems should be to help candidates enhance their skills to obtain job offers, we provide guidelines to improve existing technical interview preparation resources and motivate the design of future systems to enhance candidates' preparation processes. 

\subsubsection{Remove financial barriers}
Most technical interview preparation resources primarily focus on practicing coding skills, providing ample programming challenges based on various data structures and algorithms for candidates to solve. Many training resources also provide social and collaborative preparation methods. However, these features are often hidden behind financial barriers that require paid accounts. For instance, LeetCode Premium costs \$35 per month~\cite{leetcode-pre}, expert feedback on IGotAnOffer costs \$149 per hour~\cite{offer}, and premium interviews for interviewing.io start at \$225 per interview~\cite{interviewingio}. Our results show most candidates are unwilling to pay and often rely on free resources, such as YouTube, LeetCode (free version), and GeeksforGeeks. Financial costs also present a barrier to learning, particularly for individuals from disadvantaged groups~\cite{pennacchia2018barriers}. Thus, reducing financial barriers to more advanced features can improve candidates' preparedness and self-efficacy for technical interview evaluations. 

\subsubsection{Provide opportunities for authentic training and spectatorship}

Our results show that most candidates do not train in authentic technical interview settings---prioritizing technical skills practice over communication and social aspects of tech interviews. However, we observed candidates who practice communication more frequently and prepare with others felt significantly more prepared for technical interviews. Thus, future systems should seek to integrate functionality for authentic interview settings, such as think-aloud practice and training with others---which we found significantly improves candidates' perceived preparedness. For example, in our study we observed candidates who practice with others reported using virtual platforms, such as Discord, which has been touted as an alternative online learning medium~\cite{arifianto2021students}. In addition, Pramp provides functionality that matches users together to complete mock interviews between peers~\cite{pramp}, promoting authentic practice and collaborative learning. Recent work has also explored using artificial intelligence and LLMs to simulate general interviews~\cite{chou2022ai,agrawal2024impact,liaw2023artificial,career_io} and technical interviews~\cite{swanand2025chase,final_round_ai,aptico} in support of candidates' preparation for job assessments. 

Another way technical interview preparation platforms can enhance candidates' preparation efforts is by fostering \textit{spectatorship}, or the act of watching others complete a task. We were surprised to find YouTube is the most frequently used resource for technical interview preparation reported by participants, contrary to prior work suggesting LeetCode is the most popular~\cite{cui2024much}. However, studies show YouTube is commonly used for learning programming concepts by diverse learners~\cite{chtouki2012impact,kadriu2020investigating}. An advantage for YouTube is that it supports spectatorship---such as watching YouTube videos of users solve programming problems, which can support learning. For instance, prior work investigates spectatorship in creative writing, and found that users watching other writers helped them understand and improve their own writing processes~\cite{WatchMeWrite}. Spectatorship of gameplay videos on YouTube enhances information sharing and learning---helping watchers adopt new techniques and strategies for their own video games~\cite{golob2021video}. Live streaming, where individuals broadcast their work to live viewers, is also an emerging practice for scaling CS education~\cite{yan} and understanding software development practices~\cite{alaboudi2023constitutes}. Incorporating these functionalities into tech interview training resources can enhance candidates' learning and preparation for job assessments.

\subsection{SE Education}\label{sec:guide-edu}

Finally, we provide guidelines to improve SE education to enhance tech interview preparation for candidates. We specifically focus on university-level computing education within academic institutions. One of the core responsibilities of universities is to shape students through the education of their chosen discipline. While several participants in our survey were not from CS backgrounds, we focus on enhancing CS and SE education as they are tasked with preparing students for computing careers and represent the majority of candidates completing technical interviews. However, due to the dynamic nature of computing, SE education requires constant adjustment to meet the evolving needs of the tech industry~\cite{2}.

\subsubsection{Integrate authentic interview training in classrooms}

SE courses are intended to provide students with the knowledge and skills to succeed as software engineers~\cite{shaw2000software}. However, only one survey participant mentioned an SE course was useful for their technical interview preparation---with most noting Data Structures and Algorithms-related courses as more helpful. Further, most participants found their undergraduate curriculum in general not useful for preparing for technical interviews. Thus, computing curricula can be enhanced to support candidates pursuing SE jobs. Prior work explores adding technical interview specific courses or interview practice into existing CS courses~\cite{kapoor2021introducing,griffin2022innovative}. While these lack the high stakes nature and anxiety-inducing nature of real job interviews~\cite{kapoor2023implementation}, they can provide realistic environments and social settings to train for interviews. 

\subsubsection{Incorporate SE-related Activities to Support Authentic Interview Concepts} In addition, SE courses in particular can provide opportunities for authentic interview practice---coding and communication in front of an audience---beyond typical data structures and algorithms-focused classes. Software development is a collaborative activity, relying on teams of developers to implement and maintain software~\cite{whitehead2007collaboration}. SE-related classes often provide opportunities for collaborative learning through group projects~\cite{coppit2005large}, which has been shown to enhance learning experiences~\cite{marques2017enhancing} and promote effective communication among students~\cite{mackellar2012case}. Prior work also suggests incorporating practices from Agile---a development processes commonly used in the tech industry, such as scrum meetings~\cite{zorzo2013using} and retrospectives~\cite{krogstie2009shared}, can enhance communication and soft skills for students in SE-related classes~\cite{d2015extreme}. 

Moreover, a close environment to authentic interview settings is pair programming, where two software engineers work together on the same machine to write code~\cite{bipp2008pair}. While pair programming lacks the evaluative nature of tech interviews, Williams \etal suggest this practice is effective in educational contexts for enhancing student learning and improving communication~\cite{williams2001support}. Further, prior work suggests incorporating pair programming in technical interviews---between candidates and interviewers---can reduce anxiety and provide insights into candidates' technical abilities and communication skills in a more relevant setting~\cite{mahnaz2022async}. 

\subsubsection{Avoid Extracurricular Technical Interview Preparation Only Focused on Coding} Our results suggest the lack of time to prepare for technical interviews leads to anxiety among candidates (see Table~\ref{tab:anxiety}). This aligns with prior work, where candidates and developers decry the amount of time and effort needed to prepare for tech interviews~\cite{mahnaz2019hiring}. We also found that increasing study efforts does not enhance candidates' preparedness (see Table~\ref{tab:rq2_prep}). Based on our findings, we \textit{discourage} implementing co- and extra-curricular activities---learning activities outside of the classroom---to support technical interview preparation. Extracurricular activities have been shown to enhance academic performance and provide additional learning opportunities for students~\cite{diaz2019extracurricular}. This has also been proposed as a method to address gaps in SE education and academia in prior work~\cite{30}. Yet, while we observed increased study time improves perceived preparedness, we also found a lack of time to complete tasks was a major cause of anxiety in tech interview preparation. For example, P35 noted the difficulties of ``\textit{balancing interview prep time with other commitments like coursework}'' led to anxiety in their preparation. Further, prior work suggests students---particularly those from minority backgrounds~\cite{lunn2021uneven}---may have difficulty balancing interview preparation with coursework, extracurricular activities, and other aspects of their lives. For instance, P70 noted they ``\textit{can’t focus [on] during work days. Can only do [preparation] on weekends}''. In addition, we observed female participants felt significantly less prepared for interviews---despite spending more hours per week (7.37) on average preparing than males (6.71). Thus, additional activities may increase anxiety or stress for candidates seeking SE positions.

Alternatively, extracurricular activities that do support technical interview preparation should include authentic preparation. For instance, clubs and study groups focused on solving LeetCode problems are popular at a wide variety of institutions.\footnote{\url{https://leetcode.com/student/}} However, our results found that candidates who report practicing more LeetCode problems per day did not feel more confident in their preparation for interviews---while candidates with more communication practice and training with others had higher perceived preparedness (see Table~\ref{tab:rq2_prep}). Thus, extracurricular clubs and organizations focused on technical interview preparation should incorporate authentic interview environments, by integrating activities such as mock interviews and promoting spectatorship. Moreover, other computing-related activities can support tech interview training and job attainment. For instance, prior work shows underrepresented students often rely on minority-focused groups, such as the National Society of Black Engineers (NSBE), for networking and interview preparation~\cite{lunn2024you}. These groups also help minority candidates combat microaggressions and enhance their computing identity and sense of belonging~\cite{lunn2024you,rincon2021latinx}. 

Other technical-focused activities have also shown to support enhancing soft skills. For instance, the goal of competitive programming is to help students learn and apply algorithms---enhancing their computational thinking, problem-solving, and programming skills~\cite{yuen2023competitive}---and Moreno \etal suggest competitive programming teams can help students practice synchronous communication and team problem-solving abilities~\cite{moreno2018competitive}. Hackathons, social coding events where participants develop new software, have also been shown to promote community-based learning~\cite{lara2016hackathons}, enhancing technical abilities and soft skills such as teamwork and communication~\cite{schulten2024we}. Enhancing opportunities for these activities can help students gain practice with communication and being in front of an audience---increasing their performance in tech interview settings.

\subsubsection{Provide a Department-Specific Study Guide}
    
We found participants believe a study guide from their university would be helpful to support tech interview preparation (see Table~\ref{tab:plan}. Thus, CS departments can also explore producing a study guide or curriculum to support students pursuing SE positions specifically. Technical interviewing guidelines exist broadly (\ie Cracking the Coding Interview~\cite{CrackingCodingInterview} and the Tech Interview Handbook~\cite{handbook}) and for specific companies (\ie Google~\cite{google} and Amazon~\cite{amazon}). However, university-specific guidelines can help align preparation efforts with specific courses to support students' career goals. For instance, Duke University published a general technical interviewing guide to provide an overview of what students should expect~\cite{duke}. This could also benefit student job placement, a key metric for schools~\cite{job}. Further, instructors play a role in students' industry and interview preparedness~\cite{johnson2020teaching}. Lunn \etal show CS instructors desired institutions to provide encouragement and information to raise awareness of technical interview practices---in addition to providing instructor training and resources~\cite{lunn2024educational}. Alternatively, learning what concepts are not covered in computing curricula can highlight gaps and motivate learning outside of the classroom~\cite{li2015makes}. 

Study plans should also be embedded within curricula---as departments vary widely in terms of available classes, topics, etc.~\cite{confidential}. For instance, data structures such as hash tables and stacks---commonly used in technical interviews~\cite{CrackingCodingInterview}---are infrequently taught and assessed in CS2-level data structures courses~\cite{simon2010making}. A department-specific study guide could also provide insights on courses to take based on students' career goals. For instance, certain departments may have specialized classes focused topics such as web development~\cite{chou2002developing}, cloud computing~\cite{chen2012introducing}, and other advanced SE-related concepts. Customized study plans can help guide students through computing curricula, providing knowledge to help them succeed in tech interviews and attain computing careers.

\section{Limitations}

There are several threats to the validity of this work. Technical interview processes and preparation vary widely on different factors---such as company, role, etc. In this work, we only focus on whiteboard-style tech interviews---commonly used to evaluate candidates for a wide variety of tech roles~\cite{whiteboard}. Future work is needed to explore preparation for other techniques---such as take-home assessments~\cite{assess}. Additionally, while we recruited a diverse sample of candidates actively preparing for tech interviews in pursuit of SE roles, our results may not generalize to all candidates on the tech job market. For instance, we primarily targeted users of two platforms---Exponent~\cite{exponent} and Pramp~\cite{pramp}. However, our results found these participants used a wide range of resources beyond these tools. Our survey excludes additional factors that may impact preparation, \ie GPA, supportive home environments, etc. However, our goal was to investigate the practices, authenticity, and educational impacts on technical interview preparation. Our survey also relies on self-reported data, such as study hours per week, number of LeetCode problems completed per day, perceived anxiety, and perceived preparedness, which may be inaccurate and biased. 

\section{Future Work}

Future work can explore specific technical interview preparation resources and practices to investigate how their usage and impact on preparedness. For example, analyzing the content of tech interview-related YouTube videos. Observational approaches can provide further details about how candidates prepare for technical interviews and the resources they use. Additionally, longitudinal studies can provide additional insights on the effects of preparation, such as the types of roles candidates interview for and whether or not they receive a job offer. Based on our findings, we plan to implement novel training resources (Section~\ref{sec:guide-tools}) and explore SE curricula changes and development (Section~\ref{sec:guide-edu}) to support authentic interview preparation for candidates joining the tech workforce.


\section{Conclusion}

Technical interviews are a socially and cognitively demanding activity, leading to considerable preparation time and effort for aspiring software engineers. To understand how candidates prepare for technical interviews, we distributed an online survey to collect data from 131 individuals actively preparing for technical interviews in pursuit of SE-related roles. Our results show that participants use a variety of resources such as YouTube and LeetCode, but rarely practice in authentic interview settings. We also found SE education is inadequate for supporting technical interview preparation, and most participants perceive their preparation efforts cause anxiety and fail to prepare them for actual interviews. Based on our findings, we provide implications for enhancing technical interview preparation---providing guidelines for candidates, employers, interview prep materials, and SE education to improve candidates' preparedness for the complexity of tech hiring processes.


\section{Acknowledgments}

We would like to thank Stephen Cognetta and Lindsey Parker at Exponent/Pramp for their insights and help distributing our survey. This work was supported by a Google Award for Inclusion Research.

\balance

\bibliography{main}
\bibliographystyle{ACM-Reference-Format}

\end{document}